\documentclass[times,sort&compress,3p]{elsarticle}
\usepackage{graphicx,bm}
\usepackage{mathrsfs,amsmath,amsfonts,amssymb}
\usepackage{multirow}
\usepackage{natbib}
\usepackage{color}
\usepackage{tcolorbox}
\usepackage[utf8x]{inputenc}
\usepackage{graphicx,subfigure}
\usepackage{bbm}
\usepackage{algorithmic}
\usepackage{algorithm,float}
\newcommand{\indep}{\perp \!\!\! \perp}
\newtheorem{theorem}{Theorem}
\newtheorem{lemma}[theorem]{Lemma}
\makeatletter
\newenvironment{breakablealgorithm}
  {
   \begin{center}
     \refstepcounter{algorithm}
     \hrule height.8pt depth0pt \kern2pt
     \renewcommand{\caption}[2][\relax]{
       {\raggedright\textbf{\ALG@name~\thealgorithm} ##2\par}%
       \ifx\relax##1\relax 
         \addcontentsline{loa}{algorithm}{\protect\numberline{\thealgorithm}##2}%
       \else 
         \addcontentsline{loa}{algorithm}{\protect\numberline{\thealgorithm}##1}%
       \fi
       \kern2pt\hrule\kern2pt
     }
  }{
     \kern2pt\hrule\relax
   \end{center}
  }
\makeatother

\numberwithin{equation}{section}
\begin{document}

\begin{frontmatter}

\title{Online Sparse Sliced Inverse Regression}
\author{Haoyang Cheng, Wenquan Cui and Jianjun Xu\\\bigskip Department of Statistics and Finance\\The School of Management \\University of Science and Technology of China}
\date{\today}

\begin{abstract}
Due to the demand for tackling the problem of streaming data with high dimensional covarites, we propose an online sparse sliced inverse regression (OSSIR) method for online sufficient dimension reduction. The existing online sufficient dimension reduction methods focus on the case when the dimension $p$ is small. In this article we show that our method can achieve a better statistical accuracy and computation speed when the dimension $p$ is large. There are two important steps in our method, one is to extend the online principal component analysis to iteratively obtain the eigenvalues and eigenvectors of the kernel matrix, the other is to use the truncated gradient to achieve online $L_{1}$ regularization. We also analyse the convergence of the extended Candid covariance-free incremental PCA(CCIPCA) and our method. By comparing with several existing methods in the simulations and real data applications, we demonstrate the effectiveness and efficiency of our method.
\end{abstract}


\end{frontmatter}

\section{Introduction}
Sufficient dimension reduction (SDR) is an important branch of dimension reduction method. The assumption of SDR is that the response variable only relates to a few linear combinations of covariates, i.e
\begin{equation}\label{02}
  y \indep \bm{x} \mid\left(\bm{\beta}_{1}^{\mathrm{T}} \bm{x}, \ldots, \bm{\beta}_{d}^{\mathrm{T}} \bm{x}\right),
\end{equation}
where the response variable $y \in \mathbb{R}$ and covariates $\bm{x} \in \mathbb{R}^{p\times1}$, the directions $\bm{B} = (\bm{\beta}_{1}, \ldots, \bm{\beta}_{d}) \in \mathbb{R}^{p \times d}$ and $\indep$ denotes the independence of two variables. The column subspace of $\bm{B}$, denoted by $S(\bm{B})$, is called a SDR subspace. As the dimension reduction directions $\bm{B}$ that satisfies (\ref{02}) is not unique, we usually consider the intersection of all SDR subspace, which is denoted as $S_{y|\bm{x}}$. Li \cite{Li1991} proposed sliced inverse regression (SIR) to estimate the $S_{y|\bm{x}}$, which is an innovative method for SDR and has been widely applied in various fields. For more approaches to estimate the sufficient dimension reduction space $S_{y|\bm{x}}$, see\cite{Cook1991,zhu1996asymptotics,LiandWang2007,Li1992,Xia2002,Xia2007,MaZhu12}. However, these methods usually work well when $p$ is moderately large. It is necessary to consider new approaches to SDR when $p$ is very large or even greater than $n$. Lin\cite{LinZhaoLiu2019} proposed LassoSIR to obtain a sparse estimator of SDR directions, which is an efficient method to tackle the above problem. Besides LassoSIR, other sparse SDR methods can refer to \cite{li2006sparse, lin2017optimality, lin2018consistency}.

The previously mentioned SDR and sparse SDR methods are all designed for batch learning. However, the SDR methods for batch learning are not useful when the high dimensional data arrives sequentially. Hence, new dimension reduction methods for online learning are needed.
In batch learning, principal component analysis (PCA) and linear discriminant analysis (LDA) are two popular dimension reduction methods.
There are some PCA-based and LDA-based online dimension reduction methods, such as Incremental PCA (IPCA), see \cite{hall1998incremental,hall2000merging,weng2003candid,zhao2006novel,rodriguez2014matlab} and Incremental LDA, see \cite{pang2005incremental,zhao2008incremental,kim2007incremental,chu2015incremental}. Furthermore, the perturbation method, gradient descent optimization, and the randomized methods are widely used in the online PCA, see \cite{warmuth2008randomized,boutsidis2014online}. Besides of the above methods, there has been several studies on online SDR. Chavent\cite{Chavent2014sliced} provided a method to estimate the central dimension reduction subspace block-by-block. While considering the situation where observations arrive one-by-one, Cai\cite{Cai2020} proposed two online SDR methods based on perturbation method and gradient descent optimization. Zhang\cite{zhang2018online} adopted the idea of IPCA and extended IPCA to incremental sliced inverse regression for online SDR. As far as we know, there has been no research about online sparse SDR, only traditional SDR methods has been extended to the online methods and these methods focus on the situation when $p$ is moderately large. Hence, it is necessary to propose a new method for online sparse SDR.

In this article, we extend the Lasso-SIR to tackle the problem of online sparse SDR. In Lasso-SIR \cite{LinZhaoLiu2019}, an artificial response variable $\tilde{Y}$ is firstly constructed by the top-$d$ eigenvalues and corresponding eigenvectors of matrix $\widehat{\operatorname{cov}}(E(\bm{x}|y))$, which is the estimate of the conditional covariance matrix $\operatorname{cov}(E(\bm{x}|y))$. Then lasso method is applied to obtain an estimation of the SDR direction $\bm{\beta}$. The problem can be formulated as
\begin{equation}\label{03}
  \min_{\bm{\beta}} \frac{1}{2n}\|\tilde{Y} - \bm{X}^{T}\bm{\beta}\|_{2}^{2} + \mu\|\bm{\beta}\|_{1},
\end{equation}
where $\bm{X}$ is the $p \times n$ covariate matrix
formed by the $n$ samples, $\mu$ is the tunning parameter, $\|\cdot\|_{1}$ and $\|\cdot\|_{2}$ represent the $L_{1}$ and $L_{2}$ norm. To implement this Lasso-SIR method in an online fashion, we need to
modify several steps in the procedure of Lasso-SIR. Firstly, when the observations arrive sequentially in the data stream, which are denoted by $\left\{\left(\bm{x}_{t}, y_{t}\right), t=1,\ldots\right\}$, the update for the matrix $\widehat{\operatorname{cov}}(E(\bm{x}|y))$ will be computational costly. To reduce the computational cost, we replace the $\widehat{\operatorname{cov}}(E(\bm{x}|y))$ with another kernel matrix, see \cite{Cai2020}. Secondly, to iteratively obtain the eigenvalues and eigenvectors of the kernel matrix, we consider the online PCA method. However, in our method, to update the kernel matrix, the observations in the $t$-th step are also used in the $(t+1)$-th step, which violates the independence assumption in the online PCA. Hence, it is necessary to extend online PCA to solve this problem. Because the Candid covariance-free incremental PCA(CCIPCA) offers a good compromise between statistical accuracy and computational speed, see \cite{cardot2018online}. In this article, we choose to extend CCIPCA and derive the theoretical convergence of the extended CCIPCA. Furthermore, for the comparison, the extension of other online PCA methods are also presented, such as perturbation methods, stochastic optimization and IPCA. Thirdly, to obtain a sparse estimator in online learning, we apply the truncated gradient in our method, which has been shown to be an online counterpart of $L_{1}$ regularization in the batch setting, see \cite{John2009}. Finally, combining the above three steps, we can get the online fashion of (\ref{03}). Further, the convergence property of our OSSIR estimator is also presented. By conducting several simulations and real data analysis, we show that our method can achieve a better statistical accuracy and computation speed when dimension $p$ is large.

There are two major contributions of our method, one is that we first propose a method for online sparse SDR, the other is the avoidance of the update for the inverse of matrix covariance in every step. To our knowledge, methods in Cai\cite{Cai2020} and Zhang\cite{zhang2018online} are designed for the situation when $p$ is small. In the update step of these two methods, the computation complexity of the update for the inverse of matrix covariance is $O(p^2)$, which can not be tolerated when $p$ is large.

The outline of this article is the following. In section 2, we present the derivation of online sparse sliced inverse regression and its detail algorithm. In section 3, we demonstrate the convergence property of the extended CCIPCA and OSSIR estimator. Numerical simulations and real data applications are shown in Section 4 and 5. The article is finished with a brief conclusion in Section 6.

\section{Sparse Online Sliced Inverse Regression}
We first give a brief introduction to the Lasso-SIR\cite{LinZhaoLiu2019}. Given the samples $\left\{(y_i, \bm{x}_i)\right\}_{i=1}^{n}$, Lasso-SIR first arranges the $\{(y_i, \bm{x}_i)\}_{i=1}^{n}$ by $y_{1} \le y_{2} \le \ldots \le y_{n}$ and divides the data into $H$ equal-sized slices $I_{1}, \ldots, I_{H}$ according to $y_{i}, i = 1,\ldots,n$. For simplified, they assume $n = cH$, $E(\bm{x}) = 0$ and re-express the data as $\bm{x}_{h,j}$ and $y_{h,j}$, where $h$ and $j$ represent the slice number and the order number of a sample in the $h$th slice respectively. Then the estimation of $\bm{\Gamma} \triangleq \operatorname{cov}(E(\bm{x}|y))$ can be formulated as
\begin{equation*}
  \widehat{\bm{\Gamma}}_{H} = \frac{1}{H}\bm{X}_{H}\bm{X}_{H}^{T},
\end{equation*}
where $\bm{X}_{H} = (\bar{\bm{x}}_{1,\cdot}, \ldots, \bar{\bm{x}}_{H,\cdot})$ is a $p \times H$ matrix, $\bar{\bm{x}}_{h,\cdot}$ is the sample mean of the $h$th slice. By constructing an $n \times H$ matrix $\bm{M} = \mathbf{I}_{H} \otimes \mathbf{1}_{c}$, where $\otimes$ represents the outer product of two matrix, $\mathbf{1}_{c}$ is the $c \times 1$ vector with all entries being 1, we rewrite $\bm{X}_{H} = \bm{XM}/c$. Let $\bm{\Lambda}_{d} = diag(\hat{\lambda}_{1}, \ldots, \hat{\lambda}_{d})$ as the d-top eigenvalues of $\widehat{\bm{\Gamma}}_{H}$ and $\hat{\bm{\eta}} = \left(\hat{\bm{\eta}}_{1}, \ldots, \hat{\bm{\eta}}_{d}\right)$ as the corresponding eigenvectors, we have that
\begin{equation*}
  \widehat{\bm{\Gamma}}_{H}\hat{\bm{\eta}} =\frac{1}{H}\bm{X}_{H}\bm{X}_{H}^{T}\hat{\bm{\eta}} = \frac{1}{nc}\bm{X}\bm{M}\bm{M}^{T}\bm{X}^{T}\hat{\bm{\eta}} = \hat{\bm{\eta}}\bm{\Lambda}_{d}.
\end{equation*}
Then by setting
\begin{equation*}
  \tilde{\bm{Y}} = \frac{1}{c}\bm{M}\bm{M}^{T}\bm{X}^{T}\hat{\bm{\eta}}\bm{\Lambda}_{d}^{-1},
\end{equation*}
we have $\hat{\bm{\eta}} = \frac{1}{n}\bm{X}\tilde{\bm{Y}}$. Moreover, Li\cite{Li1991} has shown that if $\bm{x}$'s distribution is elliptically symmetric, then
\begin{equation}\label{04}
  \bm{\Sigma} \operatorname{col}(\bm{B}) = \operatorname{col}(\bm{\Gamma}),
\end{equation}
where $\bm{\Sigma}$ is the covariance matrix of $\bm{x}$, $\operatorname{col}(\bm{B})$ and $\operatorname{col}(\bm{\Gamma})$ is the space spanned by the columns of $\bm{B}$ and $\bm{\Gamma}$ respectively. With the equation (\ref{04}), we have $\bm{\eta} \propto \bm{\Sigma}\bm{\beta}$, where $\bm{\eta}$ is the eigenvector associated with the top-$d$ eigenvalues of $\bm{\Gamma}$. If we approximate $\bm{\eta}$ and $\bm{\Sigma}$ by $\hat{\bm{\eta}}$ and $\frac{1}{n}\bm{XX}^{T}$, we can obtain that
\begin{equation*}
  \frac{1}{n}\bm{X}\tilde{\bm{Y}} \propto \frac{1}{n}\bm{X}\bm{X}^{T}\bm{\beta}.
\end{equation*}
To recover a sparse vector $\hat{\bm{\beta}} \propto \bm{\beta}$, Lin \cite{LinZhaoLiu2019} consider the following optimization problem
\begin{equation}\label{06}
  \hat{\bm{\beta}}_{i} = \arg\min \mathcal{L}_{\bm{\beta},i} = \arg\min_{\bm{\beta}} \frac{1}{2n}\|\tilde{\bm{Y}}_{\star,i} - \bm{X}^{T}\bm{\beta}\|_{2}^{2} + \mu_{i}\|\bm{\beta}\|_{1}, \quad i=1,\ldots, d,
\end{equation}
where $\tilde{\bm{Y}}_{\star,i}$ is the $i$th column of $\tilde{\bm{Y}}$, $\hat{\bm{B}} = \left(\hat{\bm{\beta}}_{1}, \ldots, \hat{\bm{\beta}}_{d}\right)$ and $\mu_{i} = C\sqrt{\frac{\log(p)}{n\hat{\lambda}_{i}}}$ for sufficiently large constant C.

To turn Lasso-SIR into an online learning method, there are several steps in the procedure of Lasso-SIR that we should modify. The first is the update for the matrix $\widehat{\bm{\Gamma}}_{H}$. With streaming data $\left\{\left(\bm{x}_{i}, y_{i}\right), i=1,\ldots\right\}$, we pre-specify the cutting points $-\infty = q_{0} < q_{1} < \ldots < q_{H} = \infty$ by a small batch data and the matrix $\widehat{\bm{\Gamma}}_{H,t}$ in the $t$th step can be constructed as
\begin{equation}\label{06.1}
  \widehat{\bm{\Gamma}}_{H,t} =  \frac{1}{H}\sum_{h=1}^{H}\widetilde{\bm{X}}_{t}M_{t}M_{t}^{T}\widetilde{\bm{X}}_{t}^{T},
\end{equation}
where $\widetilde{\bm{X}}_{t} = \bm{x}_{(1:t)} - \bar{\bm{x}}_{t}\bm{1}_{t}^{T} \in \mathbb{R}^{p \times t}$, $\bm{x}_{(1:t)}$ is a $p \times t$ matrix formed by the first $t$ observations $\{\bm{x}_{i}\}_{i=1}^{t}$ and $\bar{\bm{x}}_{t} = \frac{1}{t}\sum_{i=1}^{t}\bm{x}_{i}$. $M_{t}$ is a $t \times H$ matrix and its $h$th column $\bm{M}_{t,h} = \frac{1}{n_{h}}\left(\mathbbm{1}\{y_{1} \in I_{h}\}, \ldots, \mathbbm{1}\{y_{t} \in I_{h}\}\right)^{T} \in \mathbb{R}^{t \times 1}$, $n_h$ is the sample size in the $h$th slice $I_{h}$. As we pre-specify the cutting points, the sample size in each interval usually will be different and even extreme imbalanced with the observations arrive sequentially. Therefore, \cite{Cai2020} recommended the kernel matrix in the cumulative slicing estimation\cite{Zhu2010cumulative}. Motivated by the cumulative slicing estimation, we replace $\widehat{\bm{\Gamma}}_{H}$ by the following matrix $\widehat{\bm{D}}$ in our method. Defining
\begin{equation*}
  d_{h} \triangleq E\left\{(\bm{x}-E\bm{x})I\{Y \in I_{h}\}\right\} \quad \text{and} \quad \bm{D} = \frac{1}{H}\sum_{h=1}^{H}d_{h}d_{h}^{T},
\end{equation*}
then we have $\bm{\Sigma}^{-1}\bm{D} \subset S_{y|\bm{x}}$, see Theorem 1 in \cite{Zhu2010cumulative}. Further, the matrix $\hat{d}_{t,h}$ and $\widehat{\bm{D}}_{t}$ in the $t$th step can be formulated as
\begin{equation}\label{06.2}
  \hat{d}_{t,h}  = \frac{1}{t}\sum_{i=1}^{t}(\bm{x}_{i} - \bar{\bm{x}}_{t})\mathbbm{1}\{y_{i} \in I_{h}\} \quad \text{and} \quad \widehat{\bm{D}}_{t} = \frac{1}{H}\sum_{h=1}^{H}\hat{d}_{t,h}\hat{d}_{t,h}^{T} = \frac{1}{Ht^{2}}\widetilde{\bm{X}}_{t}\widetilde{\bm{M}}_{t}\widetilde{\bm{M}}_{t}^{T}\widetilde{\bm{X}}_{t}^{T},
\end{equation}
Where $\widetilde{\bm{M}}_{t} = (\widetilde{\bm{M}}_{t,1}, \ldots, \widetilde{\bm{M}}_{t,H})$ and $\widetilde{\bm{M}}_{t,h} = \left(\mathbbm{1}\{y_{1} \in I_{h}\}, \ldots, \mathbbm{1}\{y_{t} \in I_{h}\}\right)^{T}$. To update $(\hat{d}_{t+1,1}, \ldots, \hat{d}_{t+1,H})$ from $(\hat{d}_{t,1}, \ldots, \hat{d}_{t,H})$ more efficiently, we define $\tilde{\bm{e}}_{t} = \left\{\mathbbm{1}\{y_{t} \in I_{1}\}, \ldots, \mathbbm{1}\{y_{t} \in I_{H}\}\right\}^{T} \in \mathbb{R}^{H \times 1}$, then we have
\begin{equation}\label{07}
  \hat{\bm{d}}_{t+1} = (\hat{d}_{t+1,1}, \ldots, \hat{d}_{t+1,H}) = \frac{1}{t+1}\left(\sum_{i=1}^{t}\tilde{\bm{x}}_{i}\tilde{\bm{e}}_{i}^{T} + \tilde{\bm{x}}_{t+1}\tilde{\bm{e}}_{t+1}^{T}\right).
\end{equation}
Then we have the update formula for $\hat{\bm{d}}_{t}$ and $\widehat{\bm{D}}_{t}$. With the discussion in \cite{Cai2020,Zhu2010cumulative}, we can also have that $\widehat{\bm{D}}_{t} = \bm{D} + O_{p}(t^{-1/2})$.
\subsection{Update for the Artificial Response Variable $\tilde{Y}$}
To implement (\ref{06}) in an online fashion, one of the problems we ran into is the update for $\tilde{Y}$. In this section, we will show how to update the $\tilde{Y}$ efficiently. By setting
\begin{equation*}
  \tilde{\bm{Y}}_{t} = \frac{1}{Ht}\widetilde{\bm{M}}_{t}\widetilde{\bm{M}}_{t}^{T}\bm{X}_{t}^{T}\hat{\bm{\eta}_{t}}diag(\frac{1}{\hat{\lambda}_{t,1}}, \ldots, \frac{1}{\hat{\lambda}_{t,d}}),
\end{equation*}
we find that estimating $\tilde{\bm{Y}}_{t}$ at the $t$-th iteration means to seek for the top-$d$ eigenvalues and eigenvectors of $\widehat{\bm{D}}_{t}$. To iteratively obtain the eigenvalues and eigenvectors of a covariance matrix, it is natural to consider the online PCA. There has been many approaches to online PCA, such as stochastic gradient algorithms for online PCA \cite{sanger1989optimal,krasulina1970method,oja1985stochastic,oja1992principal}, CCIPCA \cite{weng2003candid}, incremental PCA\cite{arora2012stochastic}. The computational cost and memory usage of these online PCA method in per iteration has been presented in Table 1 in \cite{cardot2018online}.

However, in the update formula of $\bm{d}_{t}$, we can find that the $\bm{d}_{t-1}$ is used in the $t$-th step update and is highly correlated with $\bm{d}_{t}$, which violates the independence assumption of online PCA. Hence, it is necessary to reformulate the iterative algorithm and analysis the convergence property of the online PCA when the independence assumption is violated. Cardot\cite{cardot2018online} has presented several online PCA methods, such as perturbation methods, incremental methods, and stochastic optimization. In these methods, CCIPCA offers a good compromise between statistical accuracy and computational speed as a method of online PCA. Thus, we choose to extend CCIPCA to perform an online eigenvalue decomposition of $\widehat{\bm{D}}_{t}$. For comparison, we will also show the extension of other three online PCA methods, such as perturbation methods, stochastic optimization and IPCA. Next, we will summarize these four different methods and analysis their computation complexity.

Firstly, the update algorithm of CCIPCA for $\widehat{\bm{D}}_{t+1}$ is as follows:
\begin{equation}\label{10}
  \begin{split}
     \bm{v}_{t+1,j} & = \frac{t}{t+1}\bm{v}_{t,j} + \frac{1}{t+1}\bm{d}_{t+1}(j)\bm{d}_{t+1}(j)^{T}\frac{\bm{v}_{t,j}}{\|\bm{v}_{t,j}\|},\\
     \bm{d}_{t+1}(j) & = \bm{d}_{t+1}(j-1) - \frac{\bm{v}_{t+1,j-1}}{\|\bm{v}_{t+1,j-1}\|}\frac{\bm{v}_{t+1,j-1}^{T}}{\|\bm{v}_{t+1,j-1}\|}\bm{d}_{t+1}(j-1),
  \end{split}
\end{equation}
where $j = 1,\ldots,d$, $\bm{d}_{t+1}(1) = \bm{d}_{t+1}$. Then the normalized eigenvector $\hat{\bm{\eta}}_{t+1,j}$ and eigenvalue $\lambda_{t+1,j}$ of $\widehat{\bm{D}}_{t+1}$ are estimated by
\begin{equation*}
  \hat{\bm{\eta}}_{t+1,j} = \bm{v}_{t+1,j}/\|\bm{v}_{t+1,j}\| \text{ and } \lambda_{t+1,j} = \|\bm{v}_{t+1,j}\|.
\end{equation*}
If $j = t$, initialize the $j$th eigenvector as $\bm{v}_{t,j} = \bm{d}_{t}(j)$. The computation complexity of (\ref{10}) is $O_{p}(pdH)$.

Secondly, the idea in \cite{Cai2020} is to use the perturbation theorem to implement online singular value decomposition. Similarly, we consider the lemma \ref{lemma01} in Appendix and have $ \overline{\bm{D}}_{t+1} = \overline{\bm{D}}_{t} - (t+1)^{-1}\left( \overline{\bm{D}}_{t} - \widehat{\bm{D}}_{t+1}\right)$ by setting $\overline{\bm{D}}_{t} = \frac{1}{t}\sum_{i=1}^{t}\widehat{\bm{D}}_{i}$. Then define $\bm{Q} = \overline{\bm{D}}_{t}$, $\bm{G} = \overline{\bm{D}} - \widehat{\bm{D}}_{t+1}$ and $\epsilon = -(t+1)^{-1}$ in Lemma \ref{lemma01}, the update algorithm of online singular value decomposition for $\widehat{\bm{D}}_{t+1}$ is as followings:
\begin{align}\label{08}
  \hat{\lambda}_{t+1,j} & = \hat{\lambda}_{t,j} - (t+1)^{-1}\hat{\bm{\eta}}_{t,j}^{T}(\overline{\bm{D}}_{t} - \widehat{\bm{D}}_{t+1})\hat{\bm{\eta}}_{t,j}, \\
  \hat{\bm{\eta}}_{t+1,j} & = \hat{\bm{\eta}}_{t,j} - (t+1)^{-1}(\hat{\lambda}_{t,j}\bm{I}_{p \times p} - \overline{\bm{D}}_{t})^{+}(\overline{\bm{D}}_{t} - \widehat{\bm{D}}_{t+1})\hat{\bm{\eta}}_{t,j}.
\end{align}

As the computation complexity of the matrix  $(\hat{\lambda}_{t,j}\bm{I}_{p \times p} - \overline{\bm{D}}_{t})^{+}$ is $O_p(p^{3})$ and the matrix multiply of $(\overline{\bm{D}}_{t} - \widehat{\bm{D}}_{t+1})\hat{\bm{\eta}}_{t,j}$ need to be executed d times, the computation complexity of (2.8) is $O_{p}(p^{3} + p^{2}d)$. Thus, when p is large, the computation cost of this method is not easy to bear.

Thirdly, besides the perturbation method, a more famous method is stochastic gradient optimization. Defining $\phi_{t,j} = \bm{d}_{t+1}^{T}\hat{\bm{\eta}}_{t,j}$, the update algorithm of stochastic gradient optimization for online PCA of $\widehat{\bm{D}}_{t}$ is as followings:
\begin{align}\label{09}
  \hat{\lambda}_{t+1,j} & = \hat{\lambda}_{t,j} + \gamma_{n}\left(\phi_{t,j}^{T}\phi_{t,j} - \hat{\lambda}_{t,j}\right), \\
  \hat{\bm{\eta}}_{t+1,j} & = \hat{\bm{\eta}}_{t,j} + \gamma_{n}\left[\bm{d}_{t+1} - \hat{\bm{\eta}}_{t,j}\phi_{t,j}^{T} - 2\sum_{i=1}^{j-1}\hat{\bm{\eta}}_{t,i}\phi_{t,i}^{T}\right]\phi_{t,j}.
\end{align}
The equation (2.10) is a first order approximation of the Gram-Schmidt orthonormalization of $\hat{\bm{\eta}}$, so we can also use Gram-Schmidt orthonormalization to replace (2.10). Because the computation complexity of Gram-Schmidt orthonormalization and (2.10) is $O_{p}(p^{2}d$) and $O_{p}(pdH)$ respectively, we recommend the equation (2.10) or perform Gram-Schmidt orthonormalization every L step in this article. The perturbation method and stochastic gradient optimization has been discussed in Cai\cite{Cai2020}, we recalculate the computation complexity of their method and obtain the computation complexity of (2.8) and (2.10) is $O_{p}(p^{3} + p^{2}d)$ and $O_{p}(pdH)$. While the computation complexity of the perturbation method and stochastic gradient optimization in Cai\cite{Cai2020} is $O_{p}(p^{3}d)$ and $O_{p}(p^{2}d)$.

Finally, Zhang\cite{zhang2018online} applied IPCA to the matrix $\Sigma^{-1/2}\Gamma\Sigma^{-1/2}$, here we use this method for the matrix $\hat{\bm{D}}_{t}$. When we have a new observation $(\bm{x}_{t+1}, y_{t+1})$, we first locate which slice it belongs to according to
the distances from $y_{t+1}$ to sample slice mean values $\bar{y}_{h}$ of the response variable. Let us suppose the
distance from $y_{t+1}$ to $\bar{y}_{k}$ is the smallest. So we place the new observation into the slice k. We denote $d_{t+1,k}$ as a new observation for $E\left\{(\bm{x}-E\bm{x})I\{Y \in I_{k}\}\right\}$, then we define a residual
\begin{equation*}
  \bm{v}_{t+1} = d_{t+1,k} - \bm{\eta}_{t}\bm{\eta}_{t}^{T}d_{t+1,k}.
\end{equation*}

Thus we have that the new $\bm{\eta}_{t+1}$ and $\Lambda_{t+1,d}$ is the top-d eigenvectors and eigenvalues of
\begin{equation}\label{11}
  \left[\bm{\eta}_{t}, \frac{\bm{v}_{t+1}}{\|\bm{v}_{t+1}\|} \right]^{T}\hat{\bm{D}}_{t+1}\left[\bm{\eta}_{t}, \frac{\bm{v}_{t+1}}{\|\bm{v}_{t+1}\|} \right] =  \left[\bm{\eta}_{t}, \frac{\bm{v}_{t+1}}{\|\bm{v}_{t+1}\|} \right]^{T}\hat{\bm{d}}_{t+1}\hat{\bm{d}}_{t+1}^{T}\left[\bm{\eta}_{t}, \frac{\bm{v}_{t+1}}{\|\bm{v}_{t+1}\|} \right].
\end{equation}
The computation complexity of (\ref{11}) and its eigen-decomposition is $O_{p}(pH(d+1) + H(d+1)^2)$ and $O_{p}((d+1)^{3})$, then the computation complexity for reduced rank incremental PCA of $\hat{\bm{D}}_{t}$ is $O_{p}\left(pH(d+1) + H(d+1)^2 + (d+1)^{3}\right)$. The computation complexity of all methods is summarized in the following table.
\begin{table}[h!t]
  \centering
  \caption{Computation complexity of online PCA for $\hat{\bm{D}}_{t}$ per iteration}
  \begin{tabular}{c|c}
    \hline
    Method & Computation Time \\
    \hline
    CCIPCA & $O_{p}(pdH)$ \\
    Perturbation & $O_{p}(p^{3} + p^{2}d)$ \\
    SGB & $O_{p}(pdH)$ or $O_{p}(p^{2}d + pdH)$ \\
    IPCA & $O_{p}\left(pH(d+1) + H(d+1)^2 + (d+1)^{3}\right)$ \\
    \hline
  \end{tabular}
\end{table}

With the above methods, we can iterative obtain the eigenvalues $\bm{\Lambda}_{t,d}$ and eigenvectors $\hat{\bm{\eta}}_{t}$ of $\hat{\bm{D}}_{t}$. Then with a new observation $(\bm{x}_{t+1}, y_{t+1})$ arriving, we construct the new artificial response $\tilde{y}_{t+1}$ by
\begin{equation}\label{13}
  \tilde{y}_{t+1} = \frac{1}{(t+1)H} \tilde{\bm{e}}_{t+1}\hat{\bm{d}}_{t+1}^{T}\hat{\bm{\eta}}_{t+1}\bm{\Lambda}_{t+1,d}^{-1}.
\end{equation}
\subsection{The Algorithm of OSSIR}
After we get the update for the artificial response variable $\tilde{Y}$, we apply the truncated gradient for least squares in \cite{John2009} to obtain the sparse estimator. The algorithm of online sparse sliced inverse regression with truncated gradient is presented as follows:
\begin{breakablealgorithm}
  \caption{Online Sliced Inverse Regression With Truncated Gradient}
  \hspace*{0.02in} \raggedright {\bf Input:} threshold $\theta \ge 0$, gravity sequence $g_{i} \ge 0$, learning rate $\gamma \in (0,1)$, $(\bm{x}_{i}, y_{i}),i=1,\ldots$                       

  \hspace*{0.02in} \raggedright {\bf Output:}
     $\hat{\bm{B}} = \left(\hat{\bm{\beta}}_{1}, \ldots, \hat{\bm{\beta}}_{q}\right)$

  \begin{algorithmic}[1]
      \STATE Initialize $\hat{\bm{D}}$ to obtain the corresponding eigenvalues and eigenvectors $(\hat{\lambda}_{1,j}, \hat{\bm{\eta}}_{1,j}), j =1, \ldots, d$ with a small batch sample $\{\bm{x}_{i}, y_{i}\}_{i=1}^{t}$.
      \FOR{i = t+1,t+2,\ldots}
      \STATE The new unlabeled example is $\bm{x}_{t+1} = [x^{1},\ldots, x^{p}]$;
      \STATE Update $\hat{\bm{d}}_{t+1}$ and $\hat{\bm{D}}_{t+1}$ by (\ref{07});
      \STATE Update $(\hat{\lambda}_{t+1,j}, \hat{\bm{\eta}}_{t+1,j}), j =1, \ldots, d$ by online PCA in section 2.1;
      \STATE Construct the new $\tilde{y}_{t+1}$ by (\ref{13});
      \FOR{j = 1, \ldots, d}
      \FOR{coefficient $\beta^{\ell}(\ell=1,\ldots,p)$}
      \STATE \textbf{if} $\beta^{\ell} > 0$ and $\beta^{\ell} \le \theta$ \textbf{then} $\beta^{\ell} \leftarrow \max\{\beta^{\ell}-g_{i}\gamma, 0\}$
      \STATE \textbf{elseif} $\beta^{\ell} < 0$ and $\beta^{\ell} \ge -\theta$ \textbf{then} $\beta^{\ell} \leftarrow \min\{\beta^{\ell}+g_{i}\gamma, 0\}$
      \ENDFOR
      \STATE Compute prediction $\hat{y} = \sum_{\ell=1}^{p}\beta^{\ell}x^{\ell}$
      \STATE Update for all $\ell$: $\beta^{\ell} \leftarrow \beta^{\ell} + 2\gamma(y-\hat{y})x^{\ell}$, $\hat{\bm{\beta}}_{t+1,j} = (\beta_{t+1,j}^{1}, \ldots, \beta_{t+1,j}^{p})$
      \ENDFOR
      \ENDFOR
  \end{algorithmic}
\end{breakablealgorithm}

The line 7 -- line 14 in Algorithm 1 is the detail steps of truncated gradient for least square. The truncated gradient method can be an online counterpart of $L_{1}$ regularization in the batch setting, see John\cite{John2009}. Hence, we can turn the (\ref{03}) to an online method by the truncated gradient. The brief description of truncated gradient is presented in Appendix.

Note that we do not need to calculate $\hat{\bm{D}}_{t}$ for all methods, only the perturbation method need the matrix $\hat{\bm{D}}_{t}$, other method only need to calculate $\hat{\bm{d}}_{t}$. Hence, if we choose the perturbation method, the computation complexity of Algorithm 1 is $O_{p}(p^{3} + p^{2}d + p^{2}H + pdH + pd)$, otherwise, the computation complexity of Algorithm 1 is $O_{p}(pdH + pd + L)$, where $O_{p}(L)$ is the computation complexity of online PCA and presented in Table 1. While Cai\cite{Cai2020} need to update the $\widehat{\Sigma}_{t}^{-1}$ and $(\hat{\bm{m}}_{t,1}, \ldots, \hat{\bm{m}}_{t,H}) = \widehat{\Sigma}_{t}^{-1}\hat{\bm{d}}_{t}$ every step, the computational cost of \cite{Cai2020} is of order $O(p^{2}d + p^{2}H + p(p+1)H + p^{2})$ or $O(p^{3}d + p^{2}H + p(p+1)H + p^{2})$. Therefore, we can conclude that our method is more computational effective than algorithm in \cite{Cai2020}.

\section{Convergency Properties}
In this section, we will discuss the some properties of our method. To analyze the convergence property of our method, we first refer to two theorems about the relationship between $L_{1}$ regularization and truncated gradient, and the consistency property of Lasso-SIR. John\cite{John2009} has analysed the relationship between $L_{1}$ regular and truncated gradient. The detail is described in Theorem \ref{thm01}.
\begin{theorem}\label{thm01}
Consider sparse online update rule (\ref{12}) with $\bm{\beta}_1 = 0$ and $\gamma > 0$. If $L(\bm{\beta},\bm{z})$ is convex in $\bm{\beta}$ and there exist non-negative constants A and B such that $\|\nabla_{1}L(\bm{\beta},\bm{z})\|^2 \le AL(\bm{\beta},\bm{z}) + B$ for all $\bm{\beta} \in R^{d}$ and $z \in R^{d+1}$, then for all $\bar{\bm{\beta}} \in R^{d}$ we have
\begin{equation}
  \begin{aligned}
    & \frac{1-0.5 A \gamma}{T} \sum_{i=1}^{T}\left[L\left(\bm{\beta}_{i}, \bm{z}_{i}\right)+\frac{g_{i}}{1-0.5 A \gamma}\left\|w_{i+1} \cdot I\left(\bm{\beta}_{i+1} \leq \theta\right)\right\|_{1}\right] \\
    \leq & \frac{\gamma}{2} B+\frac{\|\bar{\bm{\beta}}\|^{2}}{2 \gamma T}+\frac{1}{T} \sum_{i=1}^{T}\left[L\left(\bar{\bm{\beta}}, \bm{z}_{i}\right)+g_{i}\left\|\bar{\bm{\beta}} \cdot I\left(\bm{\beta}_{i+1} \leq \theta\right)\right\|_{1}\right],
  \end{aligned}
\end{equation}
where for vectors $v = [v_1,\ldots,v_d]$ and $v^{\prime} = [v_{1}^{\prime},\ldots,v_{d}^{\prime}]$, we let
\begin{equation*}
  \left\|v \cdot I\left(\left|v^{\prime}\right| \leq \theta\right)\right\|_{1}=\sum_{j=1}^{d}\left|v_{j}\right| I\left(\left|v_{j}^{\prime}\right| \leq \theta\right),
\end{equation*}
where $I(\cdot)$ is the set indicator function.
\end{theorem}
Because the loss function in our method is square loss, then by taking $\gamma = O(1/\sqrt{T})$ and with Theorem \ref{thm01}, we have $\|\bm{\beta}_{T} - \bar{\bm{\beta}}\| = O(1/\sqrt{T})$, more details can be found in section 3.4 of \cite{John2009}.

Before present the consistency property of Lasso-SIR in \cite{LinZhaoLiu2019}, we need some following technical conditions:
\begin{itemize}
  \item[(C1)] There exist constants $C_{min}$ and $C_{max}$ such that $0 < C_{min} < \lambda_{min}(\Sigma) ≤ \lambda_{max}(\Sigma) < C_{max}$
  \item[(C2)] There exists a constant $\kappa \ge 1$, such that
      \begin{equation*}
        0<\lambda=\lambda_{d}\left(\operatorname{var}(\mathbb{E}[\boldsymbol{x} \mid y]) \leq \ldots \leq \lambda_{1}\left(\operatorname{var}(\mathbb{E}[\boldsymbol{x} \mid y]) \leq \kappa \lambda \leq \lambda_{\max }(\boldsymbol{\Sigma}) ;\right.\right.
      \end{equation*}
  \item[(C3)] The central curve $m(y) = E(\bm{x}|y)$ satisfies the sliced stability condition;
  \item[(C4)] The observations $(\bm{x}_{i}, y_{i}), i= 1, 2, \ldots$ are independent and identically distributed;
  \item[(C5)] The nonzero eigenvalues of $\bm{D}$ are all distinct;
  \item[(C6)] The tuning parameter $\gamma_{t}$ in SGD satisfies $\gamma_{t} = Ct^{-1}$ for some constant C.
\end{itemize}
Condition (C1)-(C3) is described and necessary in Lin\cite{LinZhaoLiu2019}, the others is presented in Cai\cite{Cai2020}. Then the detail of the consistency of estimator $\hat{B}$ in Lasso-SIR is as follows:
\begin{theorem}\label{thm02}
Assume that $n\lambda = p^{\alpha}$ for some $\alpha > 1/2$, where $\lambda$ is the smallest nonzero eigenvalue of $var(E[x|y])$, and that conditions (C1)-(C3) hold for the multiple
index model (\ref{02}). Assume further that the dimension d of the central subspace is known. Let $\hat{B}$ be
the output of Lasso-SIR, then
\begin{equation*}
  \|P_{\hat{B}} - P_{B}\|_{F} \le C_{1}\sqrt{\frac{s\log(p)}{n\lambda}}
\end{equation*}
holds with probability at least $1 - C_{2}\exp(-C_{3}\log(p))$ for some constants $C_{2}$ and $C_{3}$.
\end{theorem}

Then with Theorem \ref{thm01} and Theorem \ref{thm02}, the key to derive the consistency of our method is to analysis the convergence of online PCA of $\widehat{\bm{D}}_{t}$. The convergence of online principal component analysis (PCA) has been analyzed in many researches, see \cite{oja1985stochastic,arora2012stochastic,weng2003candid}   . However, the situation they consider is that the t-th data $\bm{x}_{t}$ is independent with $\bm{x}_{t-1}$ and its contribution to online covariance matrix is additive. While it is not true for the $\hat{\bm{d}}_{t}$ in our method. This makes theoretical analysis of the online PCA of $\widehat{\bm{D}}_{t}$ is more complicated. Cai\cite{Cai2020} has discussed the convergency properties of perturbation method and stochastic gradient optimization. While in our article, we show the convergence of the $\hat{\bm{\eta}}_{t}$ obtained by CCIPCA in the next Theorem \ref{thm03}.
\begin{theorem}\label{thm03}
Under Conditions (C1),(C4)-(C6), the column space of $\hat{\bm{\eta}}_{t} = (\hat{\bm{\eta}}_{t,1}, \ldots, \hat{\bm{\eta}}_{t,d})$ obtained from (\ref{10}) converges almost surely to the column space of $\Gamma_{H}$, as $t \rightarrow \infty$.
\end{theorem}
Then with Theorem 2-4, we can finally derive the consistency of our method. The proof of Theorem \ref{thm03} and Theorem \ref{thm04} is presented in Appendix.
\begin{theorem}\label{thm04}
Under Conditions (C1)-(C6), Let $\hat{\bm{B}}_{t}$ be
the output of Algorithm 1, the column space of $\hat{\bm{B}}_{t} = (\hat{\bm{B}}_{t,1}, \ldots, \hat{\bm{B}}_{t,d})$ converges almost surely to the column space of $\bm{B}$, as $t \rightarrow \infty$.
\end{theorem}

\section{Simulation}
In this section, we conduct several simulations to evaluate the performance of different methods. The data generate progress is as follows. We consider three models,
\begin{align*}
  \text{model 1: } Y & = (\beta_{1}^{T}x) + \epsilon;\\
  \text{model 2: } Y & = \sin(\beta_{2}^{T}x)\times\exp(\beta_{2}^{T}x) + \epsilon; \\
  \text{model 3: } Y & = sgn(\beta_{3}^{T}x)\times\left|2 + (\beta_{4}^{T}x)/4\right|^{3} + \epsilon.
\end{align*}
where $\bm{x}$ is generated from multivariate normal distribution with zero mean and covariance structure like
\begin{equation*}
  Cov(x_{i},x_{j}) = \rho^{|i-j|}
\end{equation*}
with $\rho = 0.3$.

For model 1-3, $\beta$ is a p-dimensional vector. We set $\beta_{1,j} = 1$ for $j = 1,2$ and $\beta_{1,j} = 0$ otherwise; $\beta_{2,j} = 1$ for $j  = 2,4,6,8,10$ and  $\beta_{2,j} = 0$ otherwise; $\beta_{3,j} = 1$ for $j = 1,2,3,4$ and $\beta_{3,j} = 0$ otherwise; $\beta_{4,j} = 1$ for $j = 5,6,7$ and $\beta_{4,j} = 0$ otherwise. For each model, we repeat our simulations $N=100$ times with samples size $n=1000$ and covariate dimension $p = 20, 100, 500, 1000$. To show the advantages of our method, we compare the following methods:
\begin{itemize}
  \item[(M1)] online sliced inverse regression via truncated gradient and perturbation method
  \item[(M2)] online sliced inverse regression via truncated gradient and gradient descent optimization
  \item[(M3)] online sliced inverse regression via truncated gradient and CCIPCA
  \item[(M4)] online sliced inverse regression via truncated gradient and incremental PCA
  \item[(M5)] Online sliced inverse regression via the perturbation method
  \item[(M6)] Online sliced inverse regression via the gradient descent optimization
  \item[(M7)] Sliced inverse regression via batch learning
  \item[(M8)] Lasso Sliced inverse regression via batch learning.
\end{itemize}

To evaluate the performance of different methods, we refer to following distance:
\begin{equation*}
  d(\beta, \hat{\beta}) = 1 - |det(\beta^{T}\hat{\beta})|,
\end{equation*}
where $det(\cdot)$ stands for the determinant operator.

The results are summarized in Table 1-2. Table 1 show the average distance between estimator $\hat{\bm{\beta}}$ and true value $\bm{\beta}_{0}$. To compare the computational efficiency of these methods, we show the averages of the computing time in Table 2. From the result, we can find that Perturbation method is not suitable for high-dimensional data due to the high computation cost. Compared with the methods in \cite{Cai2020}, we have found that our method not only cost less time, but also have a better estimation accuracy in the high dimensional data. Combining the accuracy and the computation time, we recommend the onlineLassoSIR with CCIPCA to tackle the problem of online sparse sliced inverse regression.




\begin{table}[h!t]
\setlength{\abovecaptionskip}{0.6cm}
\setlength{\belowcaptionskip}{0.3cm}

  \centering
  \caption{The averages of the distance $d(\beta, \hat{\beta})$ based on 100 replications for Model 1-3\protect\\}

  \begin{tabular}{ c |c|c|c|c|c|c|c|c|c }
      \hline
      ~ & ~ & \multicolumn{4}{c|}{O-LassoSIR} & \multicolumn{2}{c|}{O-SIR} & \multicolumn{2}{c}{~} \\
      \cline{2-10}
      ~ & p & CCIPCA & IPCA & Perturbation & GD &  Perturbation & GD & SIR & LassoSIR \\
      \hline
       \multirow{4}*{model I}  & 20 & 0.0066 & 0.0065 & 0.0066 & 0.0079 &  0.0194 & 0.0102 & 0.0065 & 0.0011 \\
      \cline{2-10}
       & 100 & 0.0091 & 0.0088 & 0.0110 & 0.0096 &  0.2654 & 0.2160 & 0.0378 & 0.0015 \\
      \cline{2-10}
      & 500 & 0.0431 & 0.0408 & 0.0433 & 0.0449 &  0.7941 & 0.7776 & 0.9710 & 0.0019 \\
      \cline{2-10}
       & 1000 & 0.1190 & 0.1105 & 0.1145 & 0.1124 &  0.9190 & 0.8904 & 0.9806 & 0.0019 \\
      \hline
      \multirow{4}*{model II}  & 20 & 0.0120 & 0.0121 & 0.0122 & 0.0121 &  0.0288 & 0.0146 & 0.0038 & 0.0023 \\
      \cline{2-10}
       & 100 & 0.0148 & 0.0187 & 0.0178 & 0.0146 &  0.2171 & 0.0190 & 0.0223 & 0.0028 \\
      \cline{2-10}
       & 500 & 0.0367 & 0.0413 & 0.0487 & 0.0357 &  0.8438 & 0.8239 & 0.9577 & 0.0027 \\
      \cline{2-10}
       & 1000 & 0.1134 & 0.1172 & 0.1220 & 0.1156 &  0.9384 & 0.9310 & 0.9773 & 0.0033 \\
      \hline
      \multirow{4}*{model III}  & 20 & 0.0565 & 0.0569 & 0.0591 & 0.0431 &  0.4775 & 0.0483 & 0.0202 & 0.0270 \\
      \cline{2-10}
       & 100 & 0.0773 & 0.0787 & 0.0787 & 0.0713 &  0.3951 & 0.4061 & 0.0815 & 0.0258 \\
      \cline{2-10}
       & 500 & 0.1363 & 0.1390 & 0.1679 & 0.1562 &  0.9747 & 0.9340 & 0.9821 & 0.0281 \\
      \cline{2-10}
       & 1000 & 0.2787 & 0.2993 & 0.3451 & 0.3100 &  0.9826 & 0.9820 & 0.9996 & 0.0261 \\
      \hline
  \end{tabular}
\end{table}

\begin{table}[h!t]
\setlength{\abovecaptionskip}{0.6cm}
\setlength{\belowcaptionskip}{0.3cm}
  \centering
  \caption{The averages of the computation time (in seconds) based on 100 replications for Model 1-3\protect\\}

  \begin{tabular}{ c|c|c|c|c|c|c|c }
      \hline
      ~ & ~ & \multicolumn{4}{c|}{O-LassoSIR} & \multicolumn{2}{c}{O-SIR}  \\
      \cline{2-8}
      ~ & p & CCIPCA & IPCA & Perturbation & GD &  Perturbation & GD \\
      \hline
       \multirow{4}*{model I}  & 20 & 0.3118 & 0.5366 & 0.8714 & 0.3317 &  0.4122 & 0.2749\\
      \cline{2-8}
       & 100 & 0.4708 & 0.7358 & 4.7487 & 0.5359 &  5.0475 & 1.062 \\
      \cline{2-8}
       & 500 & 3.797 & 5.251 & 315.3 & 5.209 &  358.0 & 84.26\\
      \cline{2-8}
       & 1000 & 20.78 & 26.64 & 3239.7 & 26.38 &  3275.5 & 817.0\\
      \hline
      \multirow{4}*{model II}  & 20 & 0.3157 & 0.5511 & 0.5417 & 0.3139 &  0.4903 & 0.2563\\
      \cline{2-8}
       & 100 & 0.4633 & 0.7272 & 3.997 & 0.4800 &  4.544 & 1.0748\\
      \cline{2-8}
       & 500 & 4.156 & 5.545 & 313.8 & 4.723 &  366.3 & 82.17 \\
      \cline{2-8}
       & 1000 & 18.87 & 24.78 & 3209.6 & 20.72 &  3254.0 & 801.77\\
      \hline
      \multirow{4}*{model III}  & 20 & 0.621 & 0.811 & 1.003 & 0.622 &  0.877 & 0.484\\
      \cline{2-8}
       & 100 & 0.934 & 1.119 & 7.928 & 0.961 &  8.249 & 1.344\\
      \cline{2-8}
       & 500 & 7.172 & 7.834 & 615.6 & 7.868 &  649.0 & 85.000\\
      \cline{2-8}
       & 1000 & 33.058 & 35.030 & 5033.6 & 36.278 &  5592.6 & 841.569\\
      \hline
  \end{tabular}
\end{table}

\section{Real data Analysis}
To further show the performance of the proposed method, we apply our method to two datasets, one is the Cpusmall dataset (http://www.cs.toronto.edu/
~delve/data/comp-activ/desc.html). This dataset contain $n=3630$ observations and $p=12$ features from a computer systems activity measures. The response variable is portion of time that cpus run in user mode. We regard this dataset as a low-dimension case regression problem. We select 1000 observations as a training set and the remaining as a test set. We choose the number of the dimension reduction directions $d = 3$. After applying the dimension reduction methods to the dataset, we use SVM algorithm to construct the regression model. We use the relative prediction error to evaluate the prediction performance, i.e $\sum_{i \in testset}(Y_{i} - \hat{Y}_{i})^2\Big/\sum_{i \in testset}(Y_{i} - \bar{Y})^2$.

The other dataset is the activity recognition based on wearable physiological measurements in this section. The dataset can be obtained from the website http://www.mdpi.com/1424-8220/19/24/5524/s1. This dataset contains $n=4480$ observations and $p=533$ features from Electrocardiogram (ECG), Thoracic Electrical Bioimpedance (TEB) and the Electrodermal Activity (EDA) for activity recognition. To be explicit, there are 174 attributes are statistics extracted from the ECG signal, 151 attributes are features extracted from the TEB signal, 104 attributes come from the EDA measured in the arm, and 104 ones from the EDA in the hand. There are four types of the activities to be analyzed, including neutral, emotional, mental and physical.
For this dataset, we still randomly select 1000 observations as a training set and the remaining as a test set. Then this dataset can be regarded as a high-dimensional classification case. We choose the number of the dimension reduction directions $d = 3$. After applying the dimension reduction methods to the dataset, we use SVM algorithm to construct the classifier. Predict accuracy
$\sum_{i \in \{testset\}} I(y_{i} = \hat{y}_{i})/n_{test}$ is used as the evaluation standards. For both dataset, LassoSIR via batch learning is regarded as a benchmark. The result is presented in the following table.
\begin{table}[h!t]
\setlength{\abovecaptionskip}{0.6cm}
\setlength{\belowcaptionskip}{0.3cm}
  \centering
  \caption{The predict accuracy in test set}
  \begin{tabular}{ |c|c|c|c|c|c|c|c| }
      \hline
      \multirow{2}*{dataset} & \multicolumn{4}{|c|}{O-LassoSIR} & \multicolumn{2}{c|}{O-SIR} & ~  \\
      \cline{2-8}
      & CCIPCA & IPCA & Perturbation & GD &  O-SIR-P & O-SIR-GD & LassoSIR \\
      \hline
      Cpusmall& 0.072 & 0.071 & 0.073 & 0.071 &  0.072 & 0.069 & 0.060\\
      \hline
      Activity Recognize & 0.640 & 0.656 & 0.611 & 0.617 &  0.402 & 0.461 & 0.689\\
      \hline
  \end{tabular}
\end{table}

From the Table 4, we can find that both online sparse SIR and online SIR\cite{Cai2020} have a similar prediction accuracy in the Cpusmall dataset. While in the activity recognition dataset, online sparse SIR have a better predict accuracy than online SIR. Moreover, compared with the benchmark, our methods are also not much inferior. Hence, it is reasonable to conclude that our method is as effective as online SIR for the low dimensional data, and more effective for the high dimensional data.

\section{Conclusion}
By implement Lasso-SIR in an online fashion, we have proposed an approach to online sparse sufficient dimension reduction, which is more computational-efficient and has a better performance than Cai\cite{Cai2020} for the high-dimensional data. Besides the update for the kernel matrix similar with \cite{Cai2020}, the online fashion of (\ref{03}) consists two important steps, one is the online update for the eigenvalues and eigenvectors of $\hat{\bm{D}}_{t}$, the other is online $L_{1}$ regularization. We slightly modify the online PCA to tackle the former problem and summarize four different methods to handle the online eigen decomposition of $\hat{\bm{D}}_{t}$. We also give the theoretical convergence of CCIPCA in our method. To tackle the sparsity problem, we use truncated gradient, which has been shown to be an online counterpart of $L_{1}$ regularization in the batch setting. Moreover, we also show the theoretical convergence properties of our estimators. From the analysis of computation complexity, we show that the computation complexity of our method is $O_{p}(pdH + pd + L)$, which is better than \cite{Cai2020} and \cite{zhang2018online}. With the simulation studies and real data analysis, we can also find that our method can achieve a better statistical accuracy and computation speed than other methods when dimension $p$ is large. However, the accuracy between our method and the batch Lasso-SIR is not enough good, which needs more further researches.
\section{Reference}
\bibliographystyle{plain}
\bibliography{ossir}

\appendix{}
\section{}
\begin{flushleft}
  \textbf{1. Perturbation
theory for online singular value decomposition}
\end{flushleft}
\begin{lemma}\label{lemma01}
  Let $\bm{Q} \in R^{p \times p}$ be a symmetric matrix and $(\lambda_j, \bm{v}_{j})$ be the eigen-pairs of $Q, j = 1, \ldots, p$. Assume $|\lambda_{1}| > \cdots > |\lambda_{d}| > \lambda_{d+1} = \cdots = \lambda_{p} = 0$. Let $\epsilon$ be a very small positive constant and $\mathbf{G}$ be a symmetric matrix. Denote the first order perturbation $\bm{Q}(\epsilon) = \bm{Q} + \epsilon\bm{G} + O(\epsilon^2)$ and the eigen-pairs of $\bm{Q}(\epsilon)$ by $\left\{\lambda_{j}(\epsilon), \bm{v}_{j}(\epsilon)\right\}$. Then
  \begin{align*}
    \lambda_{j}(\epsilon) & = \lambda_{j} + \epsilon(\bm{v}_{j}^{T}\bm{G}\bm{v}_{j}) + O(\epsilon^2), \\
    \bm{v}_{j}(\epsilon) & = \bm{v}_{j} + \epsilon(\lambda_{j}\bm{I}_{p \times p} - \bm{Q})^{+}\bm{G}\bm{v}_{j} + O(\epsilon^2), j = 1, \ldots, d,
  \end{align*}
  where $(\lambda_{j}\bm{I}_{p \times p} - \bm{Q})^{+}$ stands for the Moore-Penrose pseudo-inverse of $(\lambda_{j}\bm{I}_{p \times p} - \bm{Q})$ and $\bm{I}_{p \times p}$ stands for the $p \times p$ identity matrix.
\end{lemma}

\begin{flushleft}
  \textbf{2. The formulation of truncated gradient}
\end{flushleft}
The truncated gradient in the $t$-th step can be formulated as
\begin{equation}\label{12}
  f(\bm{\beta}_{t}) = T\left(\bm{\beta}_{t}-\gamma\nabla_{1}L(\bm{\beta}_{t}, \bm{z}_t), \lambda g_{t}, \theta\right),
\end{equation}
where $L(\bm{\beta},\bm{z})$ is the loss function, $\nabla_{1}(\bm{\beta},\bm{z})$ is a sub-gradient of $L(\bm{\beta},\bm{z})$ with respect to the first variable $\bm{\beta}$. The observation $\bm{z}_{t} = (\bm{x}_{t}, y_{t})$, $\gamma$ is the learning rate, $g_{i} > 0$ and function $T_{1}$ is defined by
\begin{equation*}
  T(v, \alpha, \theta) =
  \begin{cases}
    \max(0, v-\alpha), & \mbox{if } v \in [0,\theta] \\
    \min(0, v+\alpha), & \mbox{if } v \in [-\theta,0] \\
    v, & \mbox{otherwise}.
  \end{cases}
\end{equation*}
In the update progress, the truncated gradient can be executed every $L$ steps. If $t/L$ is not an integer, we set $g_{t}=0$; if $t/L$ is an integer, we set $g_{t}=Lg$ for a scalar $g > 0$. The larger the parameters $g$ and $\theta$ are, the more sparsity is incurred.

\begin{flushleft}
  \textbf{3. Proof of Theorem \ref{thm03}}
\end{flushleft}
As the Perturbation Methods, SGD methods and Reduced rank incremental PCA has been discussed in \cite{Cai2020} and \cite{zhang2018online}. So we only discuss the convergence of the CCIPCA in our theorem.

Firstly, we consider the situation when $j = 1$ in (\ref{10}). Then we have
\begin{equation}\label{thm03_eq01}
   \bm{v}_{t+1,1} = \frac{t}{t+1}\bm{v}_{t,1} + \frac{1}{t+1}\bm{d}_{t+1}\bm{d}_{t+1}^{T}\frac{\bm{v}_{t,1}}{\|\bm{v}_{t,1}\|},
\end{equation}
which is equivalent to
\begin{equation}\label{thm03_eq02}
  \bm{v}_{t+1,1} = \bm{v}_{t,1} + \frac{1}{t+1}\left(\frac{\bm{D}_{t+1}}{\|\bm{v}_{t,1}\|} - \bm{I}\right)\bm{v}_{t,1},
\end{equation}
\begin{equation}\label{thm03_eq03}
  \bm{v}_{t+1,1} = \bm{v}_{t,1} + \frac{1}{t+1}\left(\frac{\bm{D}}{\|\bm{v}_{t,1}\|} - \bm{I}\right)\bm{v}_{t,1} + \frac{1}{t+1}\frac{\bm{D}_{t+1} - \bm{D}}{\|\bm{v}_{t,1}\|}\bm{v}_{t,1},
\end{equation}

For the identification of the notation, we denote $v_{t,1} = v_{1}(t)$ and $\bm{D}_{t} = \bm{D}(t)$ in the following discussion. Then we refer to the following lemmas,
\begin{lemma}\label{lemma02}
  Assume that a sequence of non-negative random variables $\{r_{n}\}$ satisfy $r_{n} = O_{p}(n^{-1/2})$. Then $\forall \varepsilon > 0$
  \begin{equation}\label{lemma02_eq01}
    \lim _{k \rightarrow \infty} \operatorname{Pr}\left(\sum_{n=k}^{\infty} n^{-1} r_{n}>\varepsilon\right)=0.
  \end{equation}
\end{lemma}
Lemma \ref{lemma02} has shown that the tail of sum of root-n convergent sequences converges to zero without the independence assumption. We can refer to \cite{Cai2020} for the detail proof. Then with the Lemma \ref{lemma02}, we consider the following lemma, which is a key to the proof of Theorem \ref{thm03} and can be seen as a modified result of Theorem 2.3.1 from \cite{zhang2001convergence}.
\begin{lemma}\label{lemma03}
  Let $v_{1}^{*}$ be a locally asymptotically stable (in the sense of Liapunov) solution to
  \begin{equation}\label{lemma03_eq01}
    \dot{v}_{1}=\left(\frac{D}{\left\|v_{1}\right\|}-I\right) v_{1}
  \end{equation}
  with domain of attraction $\mathcal{D}(v_{1}^{*})$. If there is a compact set $\mathcal{A} \subset \mathcal{D}(v_{1}^{*})$ such that the solution $\bm{v}_{1}(t)$ satisfies $P\left\{v_{1}(t) \in \mathcal{A}\right\}=1$, then $v_{1}(t)$ tends to $v_{1}^{*}$ almost surely.
\end{lemma}
\textbf{Proof:} To proof this lemma, we use the Theorem 2.3.1 in \cite{kushner2012stochastic}. By (\ref{thm03_eq02}), (\ref{lemma03_eq01}) and $a_{t} = \frac{1}{t}$, The Assumption A2.2.1, A2.2.2 and A2.2.3 in \cite{kushner2012stochastic} is easy to verify. Next we will show the boundedness of $v_{1}(t)$.

By the iteration equation (\ref{thm03_eq02}), we have
\begin{equation}\label{lemma03_eq02}
  \begin{split}
     \left\|v_{1}(t)\right\|^{2} & = \left\|v_{1}(t-1)\right\|^{2}+\frac{2}{t} \frac{v_{1}^{T}(t-1) D(t) v_{1}(t-1)}{\left\|v_{1}(t-1)\right\|}-\frac{2}{t} v_{1}^{T}(t-1) v_{1}(t-1)  \\
       & +\frac{1}{t^{2}} v_{1}^{T}(t-1) D^{2}(t) v_{1}(t-1)+\frac{1}{t^{2}} v_{1}^{T}(t-1) v_{1}(t-1)-\frac{2}{t^{2}} \frac{v_{1}^{T}(t-1) D(t) v_{1}(t-1)}{\left\|v_{1}(t-1)\right\|}.
  \end{split}
\end{equation}
Next, we focus on each quantity in (\ref{lemma03_eq02}). If $\lambda_{max}(D(t)) \le \frac{1}{2}\|v_{1}(t-1)\|$,
\begin{equation}\label{lemma03_eq03}
  \frac{2}{t} \frac{v_{1}^{T}(t-1) D(t) v_{1}(t-1)}{\left\|v_{1}(t-1)\right\|} < \frac{2\lambda_{max}(D(t))}{t}\|v_{1}(t-1)\| < \frac{1}{t}\|v_{1}(t-1)\|^2.
\end{equation}
Moreover, When $t$ is large enough and satisfies $t > \max\left\{2,2\lambda_{max}^2(D(t))\right\}$,
\begin{equation}\label{lemma03_eq04}
  \frac{1}{t^2}v_{1}^{T}(t-1) D^{2}(t) v_{1}(t-1) \leq \frac{\lambda_{\max }^{2}(D(t))}{t^2} \|v_{1}(t-1)\|^2 < \frac{1}{2t}\|v_{1}(t-1)\|^2
\end{equation}
and
\begin{equation}\label{lemma03_eq05}
  \frac{1}{t^{2}} v_{1}^{T}(t-1) v_{1}(t-1) \le \frac{1}{2t}\|v_{1}(t-1)\|^2.
\end{equation}
With (\ref{lemma03_eq02}), (\ref{lemma03_eq03}), (\ref{lemma03_eq04}), (\ref{lemma03_eq05}), we have
\begin{equation}\label{lemma03_eq06}
  \begin{split}
     \left\|v_{1}(t)\right\|^{2} & < \left\|v_{1}(t-1)\right\|^{2} + \frac{1}{t}\|v_{1}(t-1)\|^2 - \frac{2}{t}\|v_{1}(t-1)\|^2 \\
      ~ & \quad + \frac{1}{2t}\|v_{1}(t-1)\|^2 + \frac{1}{2t}\|v_{1}(t-1)\|^2 - \frac{2}{t^{2}} \frac{v_{1}^{T}(t-1) D(t) v_{1}(t-1)}{\left\|v_{1}(t-1)\right\|},
  \end{split}
\end{equation}
Hence, when $t > \max\left\{2,2\lambda_{max}^2(D(t))\right\}$, we have $\left\|v_{1}(t)\right\| < \left\|v_{1}(t-1)\right\|$.

As $\|D(t) - D\| = O_p(t^{-1/2})$ and the largest eigenvalue of $D$ is bounded, when $\|v_{1}(t-1)\| < 2\lambda_{max}(D(t))$, we can also have that $v_{1}(t)$ is bounded.

Finally, from the above two cases that $\left\|v_{1}(t)\right\| < \left\|v_{1}(t-1)\right\|$ or $\|v_{1}(t-1)\| < 2\lambda_{max}(D(t))$, we can conclude that $v_{1}(t)$ is bounded with probability 1.

Besides of the boundedness of $v_{1}(t)$, we also verify the assumption A2.2.4 in \cite{kushner2012stochastic}. Define $r_{t} = \frac{\bm{D}(t+1) - \bm{D}}{\|\bm{v}_{1}(t)\|}\bm{v}_{1}(t)$, we have that
\begin{equation*}
  \|\frac{\bm{D}(t+1) - \bm{D}}{\|\bm{v}_{1}(t)\|}\bm{v}_{1}(t)\| = \|\bm{D}(t+1) - \bm{D}\| = O_p(t^{-1/2}).
\end{equation*}
Thus
\begin{equation*}
 \operatorname{Pr}\left(\sup_{m \ge k} \left\| \sum_{i=k}^{m}\frac{1}{i}\frac{\bm{D}(t+1) - \bm{D}}{\|\bm{v}_{1}(t)\|}\bm{v}_{1}(t)\right\| > \varepsilon \right) \le \operatorname{Pr}\left(\sum_{t=k}^{\infty}\frac{1}{t}r_{t} > \varepsilon \right).
\end{equation*}
With the Lemma \ref{lemma02}, the assumption A2.2.4 is satisfied. Then the Theorem 2.3.1 in \cite{kushner2012stochastic} implies the results of Lemma \ref{lemma03} here. \qed

To complete the proof of Theorem \ref{thm03}, it is necessary to show that the locally asymptotically stable solution of (\ref{lemma03_eq01}) is $\lambda_{1}\eta_{1}$ and (\ref{thm03_eq03}) satisfies $P\left\{v_{1}(t) \in \mathcal{A}\right\}=1$. Firstly, we rewrite $v_{1}(t) = \sum_{j=1}^{d}\alpha_{j}(t)\eta_{j}$, where $\alpha_{j}(t) = v_{1}^{T}(t)\eta_{j}$, $(\lambda_{j},\eta_{j}), j = 1, \ldots, d$ is the top-d eigenvalues and eigenvectors of $\bm{D}$. Then (\ref{lemma03_eq01}) is equivalent to
\begin{equation}\label{thm03_eq4}
  \begin{split}
     \dot{\alpha}^{T}\bm{\eta} & = \left(\frac{\alpha^{T}\Lambda_d\bm{\eta}}{\sqrt{\sum_{k=1}^{d} \alpha_{k}^{2}}}-\alpha^{T}\bm{\eta}\right),\\
     \dot{\alpha}  & =\left(\frac{\Lambda_d}{\sqrt{\sum_{k=1}^{d} \alpha_{k}^{2}}}-I\right) \alpha,
  \end{split}
\end{equation}
where $\alpha \triangleq  (\alpha_{1}, \ldots, \alpha_{d})$ and $\Lambda_d = diag(\lambda_{1}, \ldots, \lambda_{d})$.
Then refer to the derivation in \cite{zhang2001convergence}, we have
$\alpha_{1} \rightarrow \pm \lambda_{1}$ and $\alpha_{j} \rightarrow 0 (j > 1)$. Hence, $v_{1}(t)$ enters the domain of attraction $\mathcal{D}(\pm\lambda_{1}\eta_{1})$ with probability one. Finally, we apply Lemma \ref{lemma03} to obtain that $v_{1}(t) \rightarrow \pm\lambda_{1}\eta_{1}$ with probability 1 as $t \rightarrow \infty$. We omit the proof of the case that $j > 1$ and refer the readers to \cite{zhang2001convergence}. \qed

\begin{flushleft}
  \textbf{4. Proof of Theorem \ref{thm04}}
\end{flushleft}

For simplify, we assume $t = cH$, $E(\bm{x}) = 0$, which is same to the assumption in Lasso-SIR. Denote $\hat{\bm{\beta}}_t$ is the output of Algorithm 1. Let $\tilde{\bm{\beta}}_{lasso}$ is the solution of
  \begin{equation*}
    \min_{\bm{\beta}_{t}}\frac{1}{2t}\|\tilde{Y}_{t} - \bm{X}^{T}\bm{\beta}_{t}\|_{2}^{2} + \mu\|\bm{\beta}_{t}\|_{1},
  \end{equation*}
where the $\tilde{Y}_{t}$ is a $t \times 1$ vector, whose element is constructed from the update step of OSSIR, i.e $\tilde{Y}_{t} = \frac{1}{Ht}\widetilde{\bm{M}}_{t}\widetilde{\bm{M}}_{t}^{T}\bm{X}_{t}^{T}\hat{\bm{\eta}_{t}}diag(\frac{1}{\hat{\lambda}_{t,1}}, \ldots, \frac{1}{\hat{\lambda}_{t,d}})$. The difference with OSSIR method is that we apply lasso on the batch sample $(\tilde{Y}_{t}, \bm{X})$ here, while truncated gradient is used in OSSIR for each step. Then with Theorem 1 and the discussion in \cite{John2009}, we have that $\|\hat{\bm{\beta}}_t - \tilde{\bm{\beta}}_{lasso}\| = O_{p}(t^{-1/2})$.

Next, we set $\hat{\bm{\beta}}_{lasso}$ is the solution of
  \begin{equation*}
    \min_{\bm{\beta}}\frac{1}{2t}\|\tilde{Y} - \bm{X}^{T}\bm{\beta}\|_{2}^{2} + \mu\|\bm{\beta}\|_{1},
  \end{equation*}
where $\tilde{Y} = \frac{1}{c}\bm{M}\bm{M}^{T}\bm{X}^{T}\hat{\bm{\eta}}\bm{\Lambda}_{d}^{-1}$. Here, we can see that $\hat{\bm{\beta}}_{lasso}$ is the solution of Lasso-SIR on the first $t$ samples $(Y,\bm{X})$.

Moreover, if we arrange the $\{(y_i, \bm{x}_i)\}_{i=1}^{t}$ by $y_{1} \le y_{2} \le \ldots \le y_{t}$ and divides the data into $H$ equal-sized slices $I_{1}, \ldots, I_{H}$ according to $y_{i}, i = 1,\ldots,t$, we can find that $\bm{X} = \bm{X}_{t}$ and $\bm{M} = \widetilde{\bm{M}}_{t}$, thus $\hat{\bm{D}}_{t} = \frac{1}{H^2}\hat{\bm{\Gamma}}_{H}$.

Further, with Theorem \ref{thm03}, we have $(\hat{\lambda}_{t}(\hat{\bm{D}}_{t}), \hat{\eta}_{t}(\hat{\bm{D}}_{t}))$ converges almost surely to $(\hat{\lambda}(\hat{\bm{D}}_{t}), \hat{\eta}(\hat{\bm{D}}_{t}))$, where $(\hat{\lambda}_{t}(\hat{\bm{D}}_{t}), \hat{\eta}_{t}(\hat{\bm{D}}_{t}))$ represent the eigenvalues and eigenvectors obtained by extended online PCA; $(\hat{\lambda}(\hat{\bm{D}}_{t}), \hat{\eta}(\hat{\bm{D}}_{t}))$ represent the eigenvalues and eigenvectors obtained by SVD. With the above discussion, we can obtain that $\hat{\bm{\beta}}_{lasso}$ converges almost surely to $\tilde{\bm{\beta}}_{lasso}$.

Finally, similar with the proof of Theorem 3 in \cite{LinZhaoLiu2019}, we set $\bm{\eta}_{0} = \Sigma\bm{\beta}_{0}$, $\tilde{\bm{\eta}} = P_{\bm{\eta}_{0}}\hat{\bm{\eta}}$ and $\tilde{\bm{\beta}} = \Sigma^{-1}\tilde{\bm{\eta}} \propto \bm{\beta}_{0}$, where $\bm{\beta}_{0}$ is the true value of $\bm{\beta}$. Then we have
\begin{equation*}
  \begin{split}
     \|P_{\hat{\bm{\beta}}_{t}} - P_{\bm{\beta}_{0}}\|_{F} & = \|P_{\hat{\bm{\beta}}_{t}} - P_{\tilde{\bm{\beta}}}\|_{F} \\
       & \le 4\frac{\|\hat{\bm{\beta}}_{t} - \tilde{\bm{\beta}}\|_{2}}{\|\tilde{\bm{\beta}}\|_{2}} \\
       & \le 4\frac{\|\hat{\bm{\beta}}_{t} - \tilde{\bm{\beta}}_{lasso}\|_{2} + \|\tilde{\bm{\beta}}_{lasso} - \hat{\bm{\beta}}_{lasso}\|_{2} + \|\hat{\bm{\beta}}_{lasso} - \tilde{\bm{\beta}}\|_{2}}{\|\tilde{\bm{\beta}}\|_{2}}.\\
  \end{split}
\end{equation*}
With $\|\hat{\bm{\beta}}_t - \tilde{\bm{\beta}}_{lasso}\| = O_{p}(t^{-1/2})$, $\hat{\bm{\beta}}_{lasso}$ converges almost surely to $\tilde{\bm{\beta}}_{lasso}$ and $\frac{\|\hat{\bm{\beta}}_{t} - \tilde{\bm{\beta}}_{lasso}\|_{2}}{\|\tilde{\bm{\beta}}\|_{2}} = O_{p}(\sqrt{\frac{s\log(p)}{t\lambda}})$ obtained from Theorem 2, we have that the column space of $\hat{\bm{B}}_{t} = (\hat{\bm{B}}_{t,1}, \ldots, \hat{\bm{B}}_{t,d})$ converges almost surely to the column space of $\bm{B}$, as $t \rightarrow \infty$. \qed
\end{document}